\documentclass[aps,prd,superscriptaddress,preprintnumbers,nofootinbib,12pt]{revtex4-2}

\usepackage{graphicx,amsmath,amssymb}
\usepackage{hyperref}
\usepackage{bm}
\usepackage{hyphenat}
\usepackage{rotating}

\hypersetup{colorlinks=true, linkcolor = [rgb]{0,0.08,0.45}, citecolor = [rgb]{0,0.08,0.45}, urlcolor = [rgb]{0,0.08,0.45}}

\newcommand{\req}[1]{(\ref{#1})}
\newcommand{\lb}{\label}
\newcommand{\mbf}[1]{\mathbf{#1}}
\newcommand{\half}{\textstyle \frac{1}{2}}
\newcommand{\Scal}{\mathcal{S}}
\newcommand{\nn}{\nonumber}

\begin{document}

\preprint{SLAC-PUB-17740.}

\title{\LARGE Color symmetry and confinement as an \\  \vspace{-2pt} underlying superconformal structure \\  \vspace{-2pt} in holographic QCD \vspace{40pt}}


\author{Guy~F.~de~T\'eramond}
\email[]{guy.deteramond@ucr.ac.cr}
\affiliation{Laboratorio de F\'isica Te\'orica y Computacional, Universidad de Costa Rica, 11501 San Jos\'e, Costa Rica}

\author{Stanley~J.~Brodsky}
\email[]{sjbth@slac.stanford.edu}
\affiliation{SLAC National Accelerator Laboratory, Stanford University, Stanford, CA 94309, USA}



\begin{abstract}

\vspace{40pt}

Dedicated to the memory of our colleague, Harald Fritzsch, who, together with Murray Gell-Mann,  introduced the color quantum number as the exact symmetry responsible for the strong interaction, thus establishing quantum chromodynamics (QCD) as a fundamental non-Abelian gauge theory.   A basic understanding of hadron properties, however, such as confinement and the emergence of a mass scale, from first principles QCD has remained elusive:  Hadronic characteristics are not explicit properties of the QCD Lagrangian and perturbative QCD, so successful in the large transverse momentum domain, is not applicable at large distances.  In this article, we shall examine how this daunting obstacle is overcome in holographic QCD with the introduction of a  superconformal symmetry in  anti de Sitter (AdS) space which is responsible for confinement and the introduction of a mass scale within the superconformal group.  When mapped to light-front coordinates in physical spacetime,  this approach incorporates supersymmetric relations between the Regge trajectories of meson, baryon and tetraquark states which can be visualized in terms of specific $SU(3)_C$ color representations of quarks.  We will also briefly discuss here the implications of holographic models for QCD color transparency in view of the present experimental interest. Based on the invited contribution to the book dedicated to the memory of Harald Fritzsch.

\end{abstract}

\maketitle

\newpage

\section{Introduction}

Harald Fritzsch,  together with Murray Gell-Mann,  introduced in 1972 the color quantum number of quarks as the source of the strong interactions, with the restriction that all physical states and observables are singlets under the color group $SU(3)_C$\cite{FGM}. This  far reaching assumption provides a simple explanation of the spin-statistics problem which hindered the quark model\footnote{Previously  O. W. Greenberg proposed that quarks should be parafermions to comply with the Pauli exclusion principle~\cite{Greenberg:1964pe} and M.-Y. Han and Y. Nambu suggested the existence of three integer charged triplets with a broken $SU(3) \times SU(3)$ symmetry group~\cite{Han:1965pf}.}, but also allowed for the correct prediction of the observed decay rate of the neutral pion and the ratio $R$ for electron-positron annihilation~\cite{Bardeen:1972xk}.  While trying to understand why a single color quark or gluon cannot be observed as a physical state, known as the confinement problem,  Fritzsh and Gell-Mann proposed to use the exact color symmetry as a gauge group, thus establishing quantum chromodynamics (QCD) as a non-Abelian gauge theory describing the interactions of quark fields, in the fundamental triplet representation of the $SU(3)_C$ group, and the gluon color octet in its adjoint representation~\cite{FGM, Fritzsch:1973pi}.

The interactions between the fundamental degrees of freedom of the QCD Lagrangian, the quark and gluons, observed in high energy scattering experiments is described to high precision by QCD, establishing this theory as the standard model of the strong interactions~\cite{Gross:2022hyw}.  At low momentum transfers, however, the nonperturbative nature of the strong interactions at large distances becomes dominant, and the usual techniques of perturbative QCD become inapplicable. Despite the important advances of Euclidean lattice QCD~\cite{Wilson:1974sk} and other nonperturbative approaches,   a full understanding of color confinement, the origin of the hadron mass scale and the emergence of hadronic degrees of freedom from the  QCD Lagrangian, has remained a deep unsolved problem.

Recent theoretical developments aimed at understanding strongly coupled dynamics are based on AdS/CFT -- the correspondence between classical gravity in a higher-dimensional anti-de Sitter (AdS) space and conformal field theories (CFT) in physical space-time~\cite{Maldacena:1997re, Gubser:1998bc, Witten:1998qj}.  In practice, the AdS/CFT duality provides an effective weakly coupled classical gravity description in a ($d+1$)-dimensional AdS$_{d+1}$ space in terms of a flat $d$-dimensional strongly coupled quantum field theory defined on the AdS asymptotic boundary, the physical four-dimensional Minkowski spacetime, where the boundary conditions are imposed~\cite{Aharony:1999ti}.

Holographic models motivated by the AdS/CFT correspondence have provided a useful semiclassical approximation which captures essential features of QCD and thus can give important insights into its nonperturbative regime. Our approach to holographic QCD on the light-front (HLFQCD) is based on the embedding of Dirac’s relativistic {\it front form} of dynamics~\cite{Dirac:1949cp}  into AdS space. This precise mapping between semiclassical LF Hamiltonian equations in QCD and wave equations in AdS space~\cite{deTeramond:2008ht}, leads to relativistic boost-invariant wave equations in physical space-time, similar to the Schr\"odinger equation in atomic physics,  providing an effective computational framework for hadron properties~\cite{Brodsky:2014yha}. Further extensions of HLFQCD incorporate the exclusive-inclusive connection in QCD and provides nontrivial interconnections between the dynamics of form factors and quark and gluon distributions~\cite{deTeramond:2018ecg, Liu:2019vsn, deTeramond:2021lxc} with  pre-QCD nonperturbative approaches such as Regge theory and the Veneziano model.

In our work with Hans G\"{u}nter Dosch,  we found that a remarkable property of holographic light front QCD is the emergence of a superconformal algebraic structure~\cite{deAlfaro:1976vlx, Fubini:1984hf, Akulov:1983hjq}:  It is realized in the holographic coordinate $z$ of anti de Sitter (AdS) space and is responsible for the introduction of a mass scale within the algebra. The superconformal symmetry determines uniquely the confining interaction in the semiclassical Hamiltonian equations as well as the corresponding modification of AdS space in the infrared region. It gives rise to striking connections between the Regge trajectories of mesons, baryons and tetraquarks~\cite{Brodsky:2013ar, deTeramond:2014asa, Dosch:2015nwa} which can be visualized in terms of the fundamental $SU(3)_C$ representation of the constituent quarks as a quark-diquark cluster configuration\footnote{The idea to apply an effective  supersymmetry to hadron physics is certainly not new~\cite{Miyazawa:1966mfa, Catto:1984wi, Lichtenberg:1999sc}, but had failed to account for the special role of the pion.}. The pion is identified with the zero mode which appears in the mass spectrum of the superconformal Hamiltonian equations and has no baryonic supersymmetric partner according to the Witten index\footnote{The Witten index counts the difference between the number of bosonic and fermionic zero energy states in a supersymmetric quantum system and it is a topological invariant~\cite{Witten:1981nf}.}. This characteristic pattern is observed across all hadron families.  The resulting picture  has similarities to the duality approach described in~\cite{Fritzsch:1973pi}, where quark and gluons are not introduced initially, provided that mesons and baryons turn out to behave as if they were composite objects of quarks and gluons.

We shall give in this contribution an overview of relevant aspects of the semiclassical approximation to QCD quantized in the light front and its holographic embedding in AdS space, with an emphasis on the superconformal structure in the dual gravity theory for hadron spectroscopy and confinement. We will also discuss briefly the predictions of holographic QCD for the onset of color transparency in QCD, namely the reduced absorption of a hadron propagating in a nucleus when produced at high-momentum transfer~\cite{Brodsky:1988xz, Brodsky:2022bum}, a subject of renewed experimental interest. Other relevant aspects and applications of the light-front holographic approach to hadron physics have been described in the recent review~\cite{Gross:2022hyw}.

\section{Semiclassical approximation to light-front QCD}

A semiclassical approximation to QCD has been obtained using light-front (LF) quantization, where the initial surface is the null plane, $x^+ = x^0 + x^3 = 0$ tangent to the light cone,  thus without reference to a specific Lorentz frame~\cite{Dirac:1949cp}. Evolution in LF time $x^+$ is given by the Hamiltonian equation~\cite{Brodsky:1997de} 
\begin{align} \lb{LFHE}
i \frac{\partial}{\partial x^+} \vert \psi \rangle = P^- \vert \psi \rangle, \quad  
 P^- \vert \psi \rangle = \frac{\mbf{P}_\perp^2 + M^2}{P^+}  \vert \psi \rangle, 
\end{align}
for a hadron with 4-momentum  $P =  (P^+, P^-, \mbf{P}_{\!\perp})$, $P^\pm = P ^0 \pm P^3$, where $P^-$ is a dynamical generator and $P^+$  and $\mbf{P}_{\!\perp}$ are kinematical.   Hadron mass spectra can be computed from the LF invariant Hamiltonian 
\begin{align} \lb{P2M2}
P^2 \vert  \psi \rangle =  M^2 \vert  \psi \rangle,
\end{align}
where $P^2  =  P_\mu P^\mu = P^+ P^-  \! -  \mbf{P}_\perp^2$.  The linear evolution in Hilbert space with light-front time $x^+$ and the quantum-mechanical probabilistic interpretation of hadron states in terms of the eigenfunctions of the LF Hamiltonian~\req{P2M2} in a Fock component basis, $\vert \psi \rangle = \sum_n \psi_n  \vert n \rangle$, leads to wave equations for the light-front wave functions (LFWFs) $\psi_n$ similar to the usual Schr\"odinger equation. 

In practice, solving the actual eigenvalue problem \req{P2M2} in QCD is a formidable computational task for a non-abelian quantum field theory beyond 1 + 1 dimensions~\cite{tHooft:1974pnl}. In 1~+~1 dimensions, for example, the QCD coupling $g$ has dimensions of mass. In this case, the theory can be solved for any number of constituents and colors using discretized light-cone quantization (DLCQ) methods~\cite{Pauli:1985pv, Hornbostel:1988fb, Hornbostel:1988ne}: All physical quantities can be computed in terms of the basic 1 + 1 Lagrangian parameters, the QCD coupling and the quark masses, thus no emergent phenomena occurs. In contrast, in  3+1 physical space the coupling $g$ is dimensionless and, in the limit of massless quarks, the QCD Lagrangian is conformally invariant. Thus, in addition to the formidable computational complexity in this case, there is no indication of the origin of the hadron mass scale from the QCD Lagrangian.\footnote{Renormalization of the theory introduces a scale $\Lambda_{\rm QCD}$ which breaks conformal invariance and leads to a   logarithmic decreasing coupling $g(\mu)$, which depends  on $\Lambda_{\rm QCD}$, and to asymptotic freedom~\cite{Gross:1973id, Politzer:1973fx} for large values of  the momentum transfer  $\mu^2$.  The parameter  $\Lambda_{\rm QCD}$ is determined in high energy experiments and its value is renormalization scheme dependent. The link between the dimensionful parameter $\Lambda_{\rm QCD}$ and the dimensionless coupling $g$ is known as dimensional transmutation.}

 Starting from the QCD Lagrangian, one can express the Hamiltonian operator  $P^-$ in terms of dynamical fields, the Dirac field $\psi_+$, and the transverse gauge field $\mbf{A}$ in the $A^+ = 0$ gauge~\cite{Brodsky:1997de} 
 \begin{align}  \lb{Pm}
P^-  =  {\half} \int \! dx^- d^2 \mbf{x}_\perp \bar \psi_+ ^\dagger
\frac{ \left( i \mbf{\nabla}_{\! \perp} \right)^2 + m_q^2 }{ i \partial^+}  \psi_+
 + ~ {\rm  interactions} ,
  \end{align} 
where the interaction terms in \req{Pm} vanish in the limit $g \to 0$ and the integral is over the initial surface $x^+ = 0$, $x^\pm = x^0 \pm x^3$. For simplicity, we have omitted in \req{Pm} the contribution from the gluon field  $\mbf{A}$.

The mass spectrum is computed from the hadronic matrix element $\langle \psi(P') \vert P_\mu P^\mu  \vert\psi(P) \rangle = M^2 \langle \psi(P')   \vert\psi(P) \rangle$. For a $q \bar q$ bound state we factor out the longitudinal $X(x)$ and orbital $e^{i L \theta}$  dependence from the LFWF $\psi$  
\begin{align}
 \psi(x,\zeta, \varphi) = e^{i L \theta} X(x) \frac{\phi(\zeta)}{\sqrt{2 \pi \zeta}} ,
\end{align} 
where $\zeta^2= x(1-x) b_\perp^2$ is the invariant transverse separation between two quarks and $L$ their relative LF orbital angular momentum. The relative impact variable $\mbf{b}_\perp$ is conjugate to the relative transverse momentum $\mbf{k}_\perp$, and $x$ is the longitudinal momentum fraction $x$.  To reduce further the dynamical problem to a single variable we take the chiral limit of zero quark masses, $m_q \to 0$, to obtain~\cite{deTeramond:2008ht}
\begin{align}  \lb{M2int}
M^2  =  \int \! d\zeta \, \phi^*(\zeta) \sqrt{\zeta}
\left( -\frac{d^2}{d\zeta^2} -\frac{1}{\zeta} \frac{d}{d\zeta} + \frac{L^2}{\zeta^2}\right) \frac{\phi(\zeta)}{\sqrt{\zeta}}
+ \int \! d\zeta \, \phi^*(\zeta) U(\zeta) \phi(\zeta) ,
\end{align}
where the effective potential  $U$ has units of $M^2$ and should enforce confinement at some IR scale. It acts on the valence state and comprises all interactions, including those from higher Fock states.  The Lorentz invariant LF equation~\req{P2M2} thus becomes a LF wave equation for $\phi$~\cite{deTeramond:2008ht}
\begin{align} \lb{LFWE}
\left(-\frac{d^2}{d\zeta^2} 
- \frac{1 - 4L^2}{4\zeta^2}+ U(\zeta) \right)  \phi(\zeta) = M^2 \phi(\zeta),
\end{align}
where the critical value of the LF orbital angular momentum $L = 0$ corresponds to the lowest possible stable solution, the ground state of the light-front Hamiltonian. Eq.~\req{LFWE} is relativistic and frame-independent: It has a  structure similar to wave equations in AdS, provided that one identifies $z = \zeta$, the holographic variable~\cite{deTeramond:2008ht}.

\section{Higher spin wave equations in AdS}

Anti-de Sitter AdS$_{d+1}$ is the maximally symmetric $d+1$ space with negative constant curvature and a $d$-dimensional flat space boundary, Minkowski spacetime.  In Poincar\'e  coordinates   $x^M = \left(x^0, x^1, \cdots , x^{d - 1}, z\right)$ the asymptotic boundary  of AdS space  is given by $z = 0$.  The line element  is
\begin{align} \label{AdSm}
ds^2  &= g_{MN}dx^M dx^N  \nn \\
&= \frac{R^2}{z^2} \left(\eta_{\mu \nu} dx^\mu dx^\nu - dz^2\right),
\end{align}
where $\eta_{\mu \nu}$ is the usual  Minkowski metric in $d$ dimensions and $R$ is the AdS radius. Five-dimensional anti-de Sitter space, AdS$_5$, has 15 isometries which induce in the Minkowski spacetime boundary the symmetry under the conformal group with 15 generators in four dimensions~\cite{Aharony:1999ti}. This conformal symmetry implies that there can be no scale in the boundary theory and therefore no discrete spectrum.

The variable $z$ acts like a scaling variable in Minkowski space: different values of $z$ correspond to different energy scales at which a measurement is made.  Short space-time intervals map to the boundary in AdS space-time  near $z=0$. This corresponds to the ultraviolet (UV) region of AdS space, the conformal boundary.   On the other hand, a large  four-dimensional object of confinement dimensions  of the hadronic size maps to the large infrared~(IR) region of AdS. Thus,  in order to incorporate  confinement and discrete normalizable modes in the gravity dual, the conformal invariance must be broken by modifying  AdS space  in the IR large $z$  region,  by introducing, for example,  a sharp cut-off at the IR border,  as in the ``hard-wall’’  model of Ref.~\cite{Polchinski:2001tt},  or  by using a  ``soft-wall’’ model~\cite{Karch:2006pv}  to reproduce the observed linearity of Regge trajectories.

\subsection{Integer-spin wave equations}

The holographic embedding of the semiclassical LF bound-state wave equation  in AdS space allows us to extend \req{LFWE} to arbitrary integer spin $J$~\cite{deTeramond:2008ht, deTeramond:2013it}. To this end,  we start with the  action in AdS$_{d + 1}$ space for a tensor-$J$ field $\Phi_J = \Phi_{N_1 \dots N_J}$ in the presence of a dilaton profile $\varphi(z)$ responsible for the confinement dynamics
  \begin{align} \lb{SAdS}
  S = \int d^dx \,  dz \sqrt{g} \, e^{\varphi(z)}  \left(D_M \Phi_J D^M \Phi_J  - \mu^2 \Phi_J^2\right),
  \end{align}
 where $g$ is the determinant of the metric tensor $g_{MN}$,  $\mu$ is the AdS mass, $d$ is the number of transverse coordinates, and $D_M$ is the covariant derivative which includes the affine connection. The variation of the AdS action leads to the wave equation
  \begin{align} \lb{AdSWEJ}
      \left[
   -  \frac{ z^{d-1- 2J}}{e^{\varphi(z)}}   \partial_z \Big(\frac{e^{\varphi(z)}}{z^{d-1-2J}} \partial_z   \Big)
  +  \frac{(\mu\,R )^2}{z^2}  \right]  \Phi_J(z) = M^2 \Phi_J(z), 
  \end{align}
after a redefinition of the AdS mass $\mu$,  plus kinematical constraints which eliminate lower spin from the symmetric tensor $\Phi_{N_1 \dots N_J}$~\cite{deTeramond:2013it}. 

By substituting $\Phi_J(z) = z^{(d-1)/2 -J} e^{- \varphi(z)/2} \, \phi_J(z)$ in \req{AdSWEJ}, we find for $d = 4$ the semiclassical light-front wave equation~\req{LFWE} with 
  \begin{align} \lb{Uvarphi}
      U_J(\zeta) = \frac{1}{2} \varphi''(\zeta) + \frac{1}{4} \varphi'(\zeta)^2 + \frac{2J - 3}{2 \zeta} \varphi’(\zeta),
  \end{align} 
 as long as $z \mapsto \zeta$. This precise mapping allows us to write the LF confinement potential $U$ in terms of the dilaton profile which modifies the IR region of AdS space to incorporate confinement~\cite{Brodsky:2014yha}, while keeping the theory conformal invariant in the UV boundary of AdS  provided that  $\varphi(z) \to 0$ for $z \to 0$. The separation of kinematic and dynamic components, allows us to determine the mass  in the AdS action in terms of physical kinematic quantities with the AdS mass-radius $(\mu R)^2 =  L^2 - (2-J)^2$, consistent with the AdS stability bound~\cite{deTeramond:2008ht, deTeramond:2013it, Breitenlohner:1982jf}.

\subsection{Half-integer-spin wave equations}

A similar derivation follows from the Rarita-Schwinger action for a spinor field $\Psi_J \equiv \Psi_{N_1 \dots N_{J - 1/2}}$ in AdS$_{d + 1}$ for  half-integral spin $J$~\cite{deTeramond:2013it}.  However, in this case, the dilaton term does not lead to an interaction~\cite{Kirsch:2006he},  and an effective Yukawa-type coupling to a potential $V$ in the action has to be introduced instead~\cite{deTeramond:2013it, Abidin:2009hr, Gutsche:2011vb}:
  \begin{align} \lb{SAdSs}
  S = \int \, d^d x \,  dz \sqrt{g}  \, \overline \Psi_J  \left(i \Gamma^A e_A^M D_M  - \mu + \frac{z}{R} V(z) \right) \Psi_J,
  \end{align}
 where $e^M_A$ is the vielbein, and the covariant derivative $D_M$ on a spinor field includes the affine connection and the spin connection. The tangent space Dirac matrices, $\Gamma^A$, obey the usual anticommutation relations $\{\Gamma^A, \Gamma^B\} = 2 \eta^{A B}$.
Factoring out the four-dimensional plane-wave and spinor dependence, 
$\Psi_J(x, z)_\pm   = e^{ i P \cdot x}  z^{(d+1)/2 - J} \,   \psi_\pm(z) \, u^\pm_{\nu_1 \cdots \nu_{J-1/2}}({P})$,
we find from \req{SAdSs} the coupled linear differential equations for the chiral components $\psi_\pm$  
\begin{align} 
- \frac{d}{d z} \psi_-  - \frac{\nu+\half}{z}\psi_-  -  V(z) \psi_-&= M \psi_+ , \lb{psip} \\
 \frac{d}{d z} \psi_+ - \frac{\nu+\half}{z}\psi_+  - V(z) \psi_+ &= M \psi_-  ,  \lb{psim}
\end{align}
with $\vert\mu R \vert = \nu + \half$ and equal probability 
\begin{align}
\int d z \, \psi_+^2(z) = \int d z  \, \psi_-^2(z).
\end{align}

Eqs. \req{psip} and \req{psim} are equivalent to the second-order equations~\cite{deTeramond:2013it}
\begin{align} 
\left(-\frac{d^2}{dz^2}
- \frac{1 - 4 \nu^2}{4z^2} + U^+(z) \right) \psi_+\! & =  M^2 \psi_+ \lb{psi1} , \\
\left(-\frac{d^2}{dz^2} 
- \frac{1 - 4(\nu + 1)^2}{4z^2} + U^-(z) \right) \psi_- &=M^2 \psi_-   \lb{psi2} ,
\end{align}
with
\begin{align} \lb{UV}
U^\pm(z) = V^2(z) \pm V’(z) + \frac{1 + 2  \nu}{z} \, V(z) .
\end{align}

Embedding  semiclassical light-front Hamiltonian equations in AdS space gives important insights into the nonperturbative structure of semiclassical bound state equations in QCD for arbitrary spin, but it does not answer the question of how the effective confinement dynamics is actually determined, and how it can be related to the symmetries of QCD itself. An important clue, however, comes from the realization that the potential $V(z)$ in Eq.~\req{UV} plays the role of the superpotential in supersymmetric  quantum mechanics (QM)~\cite{Witten:1981nf}.

\section{Superconformal symmetry and emergence of a mass scale}

We follow Fubini and Rabinovici~\cite{Fubini:1984hf} and consider a one-dimensional quantum field theory invariant under conformal and supersymmetric transformations: It is a superconformal symmetry which determines uniquely the superpotential $V(z)$ and, consequently, the IR deformation of AdS space.

As a first step, we examine supersymmetric QM which is based on a graded Lie algebra consisting of two anti-commuting supercharges $Q$ and $Q^\dagger$,  $\{Q,Q\} =  \{Q^\dagger, Q^\dagger\} = 0$, which commute with the Hamiltonian $H$~\cite{Witten:1981nf}
\begin{align}
\tfrac{1}{2}   \{Q, Q^\dagger\} & =  H ,  \lb{HQQ}\\
[Q, H]  & =  [Q^\dagger, H] = 0. \lb{crQ}
\end{align}
If the state $\vert E \rangle$ is an eigenstate with energy $E$, $H\vert E \rangle = E \vert E \rangle$, then, it follows from the commutation relations \req{crQ} that the state $Q^\dagger \vert E \rangle$  is degenerate with the state $\vert E \rangle$ for $E \ne 0$, but for  $E = 0$  we have  $Q^\dagger \vert E=0 \rangle= 0$, 
namely the zero mode has no supersymmetric partner~\cite{Witten:1981nf}, a key result for deriving the supermultiplet structure and the pattern of the hadron spectrum.

As a second step, and following Fubini and Rabinovici, we consider the scale-deformed supercharge operator,
\begin{align}
R_\lambda = Q + \lambda S,
\end{align}
a superposition of supercharges within the extended graded algebra~\cite{Fubini:1984hf}, where $S$  is related to the generator of special conformal transformations $K$. The generator $R_\lambda$ is also nilpotent, $\{R_\lambda, R_\lambda\} =  \{R_\lambda^\dagger, R_\lambda^\dagger\} = 0$, and gives rise to a new scale-dependent Hamiltonian $G$, which also closes under the extended algebra and is a compact operator 
\begin{align}
\tfrac{1}{2}  \{R_\lambda, R_\lambda^\dagger\} & =  G ,  \quad  \tfrac{1}{2} \{S, S^\dagger\}  =  K, \\
[R_\lambda, G]  & =  [R_\lambda^\dagger, G] = 0.
\end{align}
The fermion supercharges $R_\lambda$ and $R^\dagger_\lambda$ have the matrix form
 \begin{align} \lb{Rex} 
 R_\lambda = 
 \left(\begin{array}{cc} 0 & r_\lambda \\ 0 & 0 \end{array}\right), \quad
R_\lambda^\dagger=\left(\begin{array}{cc} 0 & 0\\r^\dagger _\lambda&0\end{array}\right),
\end{align}
which is realized as operators in the holographic $z$-coordinate:
\begin{align}
r_\lambda = - \partial_z+\frac{f}{z}+\lambda z,  \quad \quad  r^\dagger _\lambda  = \partial_z+\frac{f}{z}+\lambda z.
\end{align}
The parameter $f$ is dimensionless and $\lambda$ has the dimension of  [$M^2$], since $z$ has dimensions [$M^{-1}$]. Therefore, a mass scale is introduced in the Hamiltonian within the conformal group. 

The Hamiltonian equation $G\vert M \rangle = M^2 \vert M \rangle$ leads to the wave equations
\begin{align} 
 \left(\! - \frac{d^2}{d z^2}  - \frac{1-  4 (f + \half)^2}{4 z^2} + \lambda ^2 \,z^2+ 2 \lambda \left( f  - \half\right) \right)  \! \phi_+ 
 = M^2  \phi_+,   \lb{phi1} ~~~~~~ \\
 \left(\! - \frac{d^2}{d z^2}  - \frac{1- 4 (f - \half)^2}{4 z^2} + \lambda^2 \, z^2 +   2 \lambda \left( f  + \half\right) \right) \! \phi_-  
 =  M^2  \phi_- ,  \lb{phi2} ~~~~~~
 \end{align}
which have the same form as the  Euler-Lagrange equations obtained from the AdS action, but here,  the interaction potential is completely fixed by the superconformal symmetry~\cite{deTeramond:2014asa, Dosch:2015nwa}.

\subsection{Light-front mapping and baryons}

Upon mapping \req{phi1} and \req{phi2} to the semiclassical LF wave equations \req{psi1} and \req{psi2} using the substitutions
$ f  \mapsto   \nu + \half, ~
 \phi_+ \mapsto \psi_- ~ {\rm and}~  \phi_- \,  \mapsto \psi_+ $, we obtain the result
 \begin{align}
 U^+(z) &= \lambda^2 z^2 + 2 \lambda(\nu + 1), \\
 U^-(z)  &=  \lambda^2 z^2 + 2 \lambda \nu.
 \end{align}
Finally, thereafter the mapping $z \mapsto \zeta$, and $\nu \mapsto L$, we find the light-front semiclassical wave equation for the components $\psi_+$ and $\psi_-$\cite{deTeramond:2014asa}
\begin{align} 
\left(-\frac{d^2}{d\zeta^2}
- \frac{1 - 4 L^2}{4\zeta^2} + \lambda^2 \zeta^2 + 2 \lambda(L + 1) \right) \psi_+\! & =  M^2 \psi_+ \lb{LFpsi1} , \\
\left(-\frac{d^2}{d\zeta^2} 
- \frac{1 - 4(L + 1)^2}{4\zeta^2} + \lambda^2 \zeta^2 + 2 \lambda L \right) \psi_- &=M^2 \psi_-   \lb{LFpsi2} ,
\end{align}
which correspond to nucleons with LF orbital angular momentum $L$ and $L + 1$. 

\begin{figure}[h]
\includegraphics[width=7.2cm]{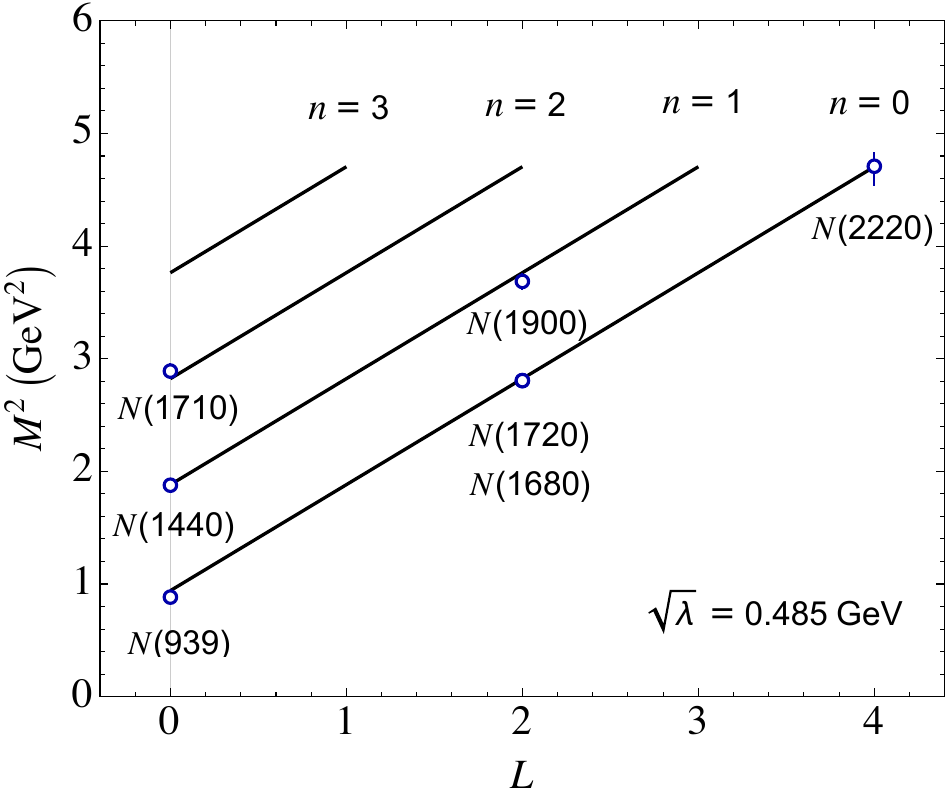} \hspace{20pt}
\includegraphics[width=7.2cm]{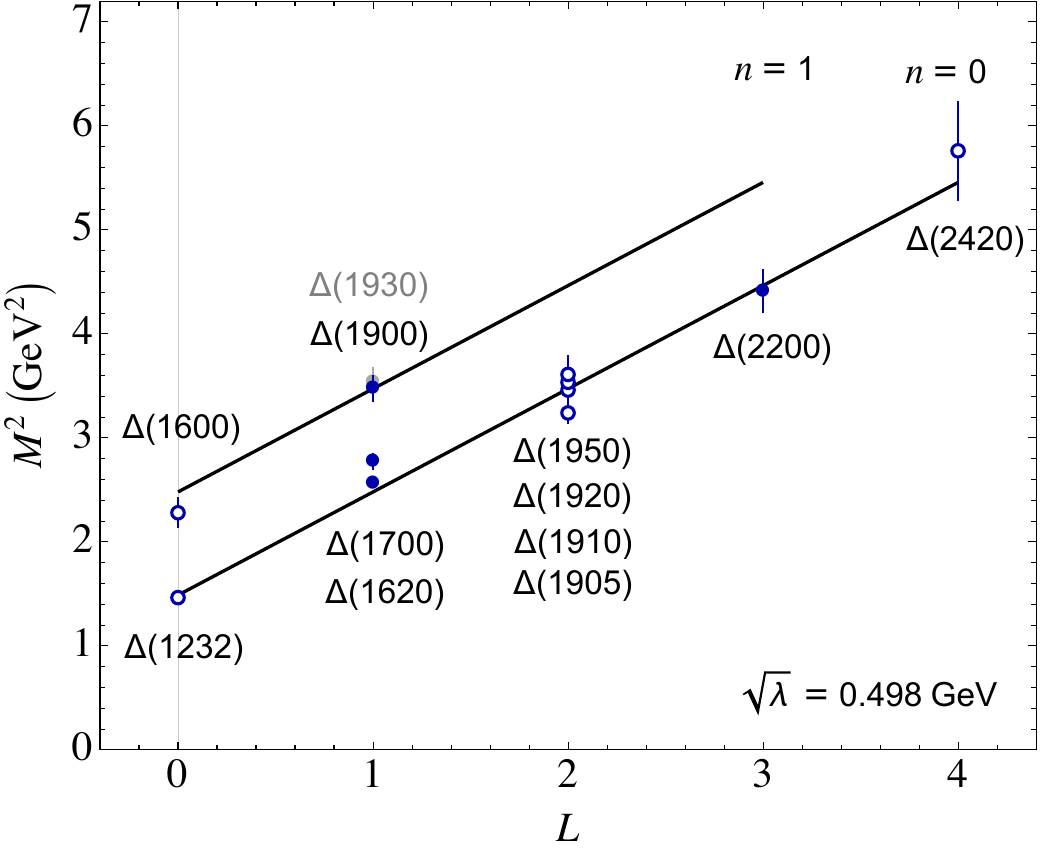}
\caption{\lb{fig:nucleon-delta}  Model predictions for the orbital and radial positive-parity nucleons (left) and positive and negative  parity $\Delta$ families  (right) compared with the data from Ref.~\cite{ParticleDataGroup:2022pth}. The values of $\sqrt{\lambda}$ are $\sqrt{\lambda} = 0.485$ GeV for nucleons and $\sqrt{\lambda} = 0.498$ GeV for the deltas.}
\end{figure}

The solutions  of \req{LFpsi1} and \req{LFpsi2}  give  the eigenfunctions  
 \begin{align}
 \psi_+(\zeta)  & \sim \zeta^{\frac{1}{2} + L} e^{-\lambda \zeta^2/2}
  L_n^L(\lambda \zeta^2), \\
\psi_-(\zeta) & \sim \zeta^{\frac{3}{2} + L} e^{-\lambda \zeta^2/2} L_n^{L+1}(\lambda \zeta^2) ,
\end{align}
 with eigenvalues  
 \begin{align}
 M^2 = 4 \lambda (n + L + 1).
 \end{align}  
 The polynomials $L_n^L(x)$ are associated Laguerre polynomials, where the radial quantum number $n$ counts the number of nodes in the wave function. We compare in Fig.~\ref{fig:nucleon-delta} the model predictions with the measured values for the positive parity nucleons~\cite{ParticleDataGroup:2022pth} for $\sqrt{\lambda} = 0.485$ GeV.  The LF semiclassical wave equations \req{LFpsi1} and \req{LFpsi2} are $J$-independent equations resulting in the absence of spin-orbit coupling, in agreement with the observed degeneracy  in the baryon spectrum  for a fixed value of $L$ along a given Regge trajectory, as illustrated in Fig.~\ref{fig:nucleon-delta}.

\subsection{\lb{scMB}Superconformal meson-baryon symmetry}

 In the usual applications of supersymmetry the supercharges connect bosonic to fermionic states. In the present holographic context superconformal quantum mechanics  leads to a relation of meson and baryon wave functions,  therefore to an effective hadronic supersymmetry which underlies the $SU(3)_C$ representation properties, since a diquark cluster can be in the same color representation as an antiquark, namely $\bf{\bar 3} \in {\bf 3} \times {\bf 3}$, thus incorporating the dynamics of color. Supersymmetry at the hadronic level is realized at larger distances, where the diquark cluster appears effectively as an antiquark. At larger momentum transfer, however, where single quarks are resolved individually, this effective supersymmetry is expected to be only an approximate symmetry. For example,  the recent analysis in Ref.~\cite{Liu:2019vsn} of the $u$-quark  distribution in the proton shows a deviation by 10--15\% from a quark-diquark configuration~\footnote{See also Sec. 5.4.13  of Ref.~\cite{Gross:2022hyw}}.

The specific meson--nucleon connection follows from the substitution $\lambda \mapsto \lambda_B=\lambda_M, \ f  \mapsto L_M-\half = L_B + \half,\ \phi_+ ~\mapsto ~\phi_M, ~ \phi_- \mapsto \phi_B$  and $z \to \zeta$ in the superconformal equations \req{phi1} and \req{phi2}. We find the LF meson (M)--baryon (B) bound-state equations~\cite{Dosch:2015nwa}
\begin{align} \lb{M}
 \left(-\frac{d^2}{d\zeta^2} - \frac{1- 4 L_M^2}{4 \zeta^2} +  \lambda_M^2\, \zeta^2 + 2 \lambda_M (L_M - 1) \right)\phi_M  &= M^2 \, \phi_M  , \\ \lb{B}
 \left(-\frac{d^2}{d\zeta^2} - \, \frac{1- 4 L_B^2}{4 \zeta^2} \, +   \lambda_B^2\, \zeta^2 + \, 2 \lambda_B (L_B +1)  \right)\phi_B &= M^2 \, \phi_B .
\end{align}

\begin{figure}[h]
\includegraphics[width=7.6cm]{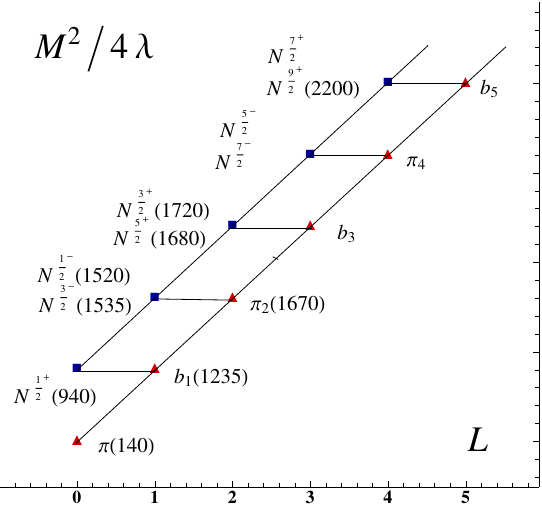}
\caption{\label{MN} This schematic figure  from Ref.~\cite{Dosch:2015nwa} illustrates the meson-nucleon superconformal connection.  The predicted value of $M^2$  in units of $4 \lambda$ for $S=0$ mesons and $S = \frac{1}{2}$ baryons is plotted vs the orbital angular momentum $L$. The $\pi$-meson has no baryonic partner.}
 \end{figure}

The superconformal structure imposes the condition  $\lambda = \lambda_M = \lambda_B$  and the remarkable relation $L_M = L_B + 1$, where $L_M$ is the  LF angular momentum between the quark and antiquark in the meson, and $L_B$ between the active quark and spectator cluster in the baryon. Likewise, the equality of the Regge slopes embodies the equivalence of the ${\bf 3} -{\bf\bar 3}$ color interaction  in the $q \bar q$ meson with the ${\bf 3} - {\bf\bar 3}$ interaction between the quark and diquark cluster in the baryon. The mass spectrum from \req{M} and \req{B} is 
\begin{align} \lb{MNspec}
M^2_M &= 4 \lambda (n+ L_M ) , \\
M^2_B &= 4 \lambda (n+ L_B+1).
\end{align}

The pion has a special role as the unique state of zero mass and, since $L_M = 0$,  the pion does not have a baryon partner, as illustrated schematically in figure~\ref{MN}.  This remarkable result underlines the dual role played by the pion in light-front holographic QCD since, as a unique state of zero energy, it formally plays the same role as the unique lowest state in a supersymmetric quantum field theory~\cite{Witten:1981nf, deAlfaro:1976vlx, Fubini:1984hf, Dosch:2015nwa}. In the light front, the supersymmetric constraint on the lowest state is realized by the exact cancellation in Eq.~\req{M} of the confinement potential and the quark LF kinetic energy for massless quarks.

\subsection{AdS warped metrics and superconformal symmetry}

Instead of introducing a dilaton profile $\varphi(z)$ in the AdS action as in Eq.~\req{SAdS} (the string frame), one can modify the the AdS metric $g_{MN}$ (the Einstein frame) by introducing  a $J$-independent warp factor:   $g_{MN} \to g_{MN} e^{2 f(z)}$.  Both descriptions are equivalent, provided that $f(z) = \varphi(z)/(d - 1)$~\cite{Brodsky:2014yha}.   The dilaton profile can be determined  by integrating Eq.~\req{Uvarphi} for a given effective potential $U_J(z)$. 
To simplify the actual computation,  we introduce the function $h(z) = e^{\varphi(z)/2}$ in~\req{Uvarphi} to obtain
\begin{align} \lb{UJh}
U_J(z) =  \frac{1}{h(z)} z^{-\gamma - 1}  \partial_z \left(z^\gamma  z \partial_z h(z) \right),
\end{align}
with $\gamma = 2J - d$: It can be expressed as the linear differential equation
\begin{align}  \lb{Uh}
\Big[z^2 \partial_z^2 - (d-1-2J ) z \partial_z - z^2  U_J(z)  \Big] h(z) = 0,
\end{align}
amenable to analytic solution.\footnote{Eq.~\req{UJh} was found previously in the context of a renormalization group flow study in Ref.~\cite{Gao:2022ojh} in terms of the one-particle irreducible effective action in AdS.}

The meson potential for $J = L$ (no internal spin) follows from~\req{M}: It is given by
\begin{align} 
U(z) =  \lambda^2 z^2 + 2 \lambda (J - 1),
\end{align}
and the corresponding solution of~\req{Uh}  by 
\begin{align}
h(z) = A\, e^{\lambda z^2/2} + B \, \Gamma\left(2-J, \lambda z^2\right) e^{-\lambda z^2/2},
\end{align}
where $\Gamma(a,b)$ is the incomplete Gamma function.  In the semiclassical approximation (neglecting backreaction from the metric), the dilaton profile should depend only on the modification of AdS space, therefore independent of $J$. This condition implies that $B = 0$, and since 
$\varphi(z) \to 0$ for $z \to 0$ we set $A = 1$. Thus the solution 
\begin{align} \lb{vphz}
    \varphi(z) = 2 \log[h(z)] = \lambda z^2,
\end{align}
imposed by the superconformal algebraic symmetry: It has the same form as the result found in Ref.~\cite{Dosch:2016zdv} using a different procedure. Since the dilaton profile~\req{vphz} is $J$-independent, an equivalent description can be given in the Einstein frame.

\subsection{\lb{si} Spin interaction and diquark clusters}

The LF bound-state  equations in AdS space allows us to extend the superconformal Hamiltonian to include the spin-spin interaction, a problem not defined in the chiral limit by standard procedures. Since the dilaton profile $\varphi(z) = \lambda z^2$ is valid for arbitrary $J$, it leads from~\req{AdSWEJ} to the additional term  $2 \lambda \Scal$ in the effective potential for mesons. We are thus led to extend the original superconformal Hamiltonian by writing
\begin{align}
G = \tfrac{1}{2} \{R_\lambda, R_\lambda^\dagger\} + 2 \lambda \Scal, 
\end{align}
which maintains the meson-baryon supersymmetry~\cite{Brodsky:2016yod}. The spin $\Scal = 0, 1$, is the total internal spin of the meson, or the spin of the diquark cluster of the baryon partner. The effect of the spin term is an overall shift of the quadratic mass, 
\begin{align}
M^2_M &= 4 \lambda (n+ L_M )  + 2 \lambda \Scal  \lb{M2S}, \\
M^2_B &= 4 \lambda (n+ L_B+1) + 2 \lambda \Scal  \lb{M2},
\end{align}
as depicted in Fig.~\ref{fig:rho-delta} for the spectra of the $\rho$ mesons and $\Delta$ baryons by shifting one unit the value of $L_B$~\cite{Dosch:2015nwa}.

\begin{figure} [h]
\includegraphics[width=7.6cm]{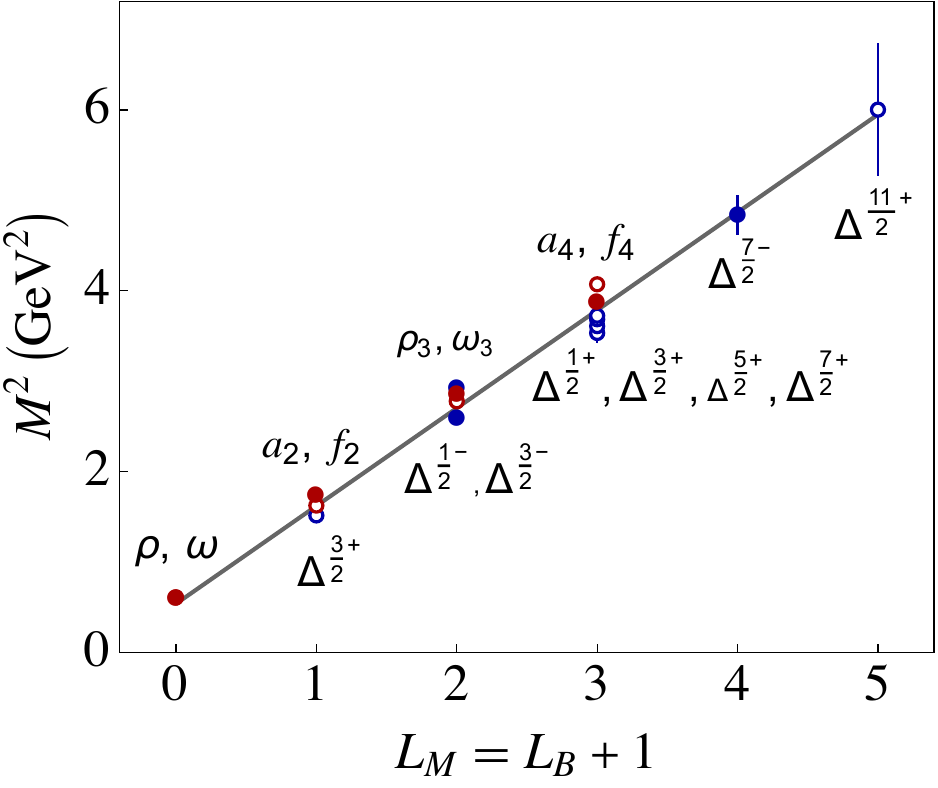}
\caption{\lb{fig:rho-delta} Supersymmetric vector meson and $\Delta$  partners from Ref.~\cite{Dosch:2015nwa}.  The experimental values of $M^2$ from Ref.~\cite{ParticleDataGroup:2022pth}  are plotted vs $L_M = L_B+1$ for  $\sqrt \lambda \simeq 0.5$ GeV. The $\rho$ and $\omega$ mesons have no baryonic partner, since it would imply a negative value of $L_B$.}
\end{figure}

For the $\Delta$ baryons the total internal spin $S$ is related to the diquark cluster spin $\Scal$ by $S = \Scal  + \frac{1}{2} (-1)^L$, and therefore, positive and negative $\Delta$ baryons have the same diquark spin, $\Scal = 1$.  As a result, all of the $\Delta$ baryons lie, for a given $n$, on the same Regge trajectory, as shown in  Fig.~\ref{fig:nucleon-delta}. Positive parity nucleons are assigned $\Scal = 0$ and are well described by the holographic model as shown in Fig.~\ref{fig:nucleon-delta}. For negative parity nucleons,  both  $\Scal = 0$  and  $\Scal = 1$ are possible, but the precise comparison with data is not as successful as for the $\Delta$ baryons and positive parity nucleons.

\subsection{Completing the supersymmetric hadron multiplet}

Besides mesons and baryons, the supersymmetric multiplet 
\begin{align}
\Phi = \begin{pmatrix}\phi_M & \phi_B^- \\ \phi_B^+& \phi_T \end{pmatrix} ,
\end{align}
contains a further bosonic partner, a tetraquark, which,  follows from the action of the SUSY operator $R^\dagger_\lambda$~\req{Rex} on the negative-chirality component of a baryon~\cite{Brodsky:2016yod}, as illustrated in Fig. \ref{fig:MBTplet}.  A clear example is the SUSY positive parity $J^P$ multiplet  $2^+,   \frac{3}{2}^+,   1^+$  of states  $f_2(1270),  ~\Delta(1232),  ~a_1(1260)$, where the $a_1$ is interpreted as a tetraquark. 

\begin{figure}[h]
\includegraphics[width=7.2cm]{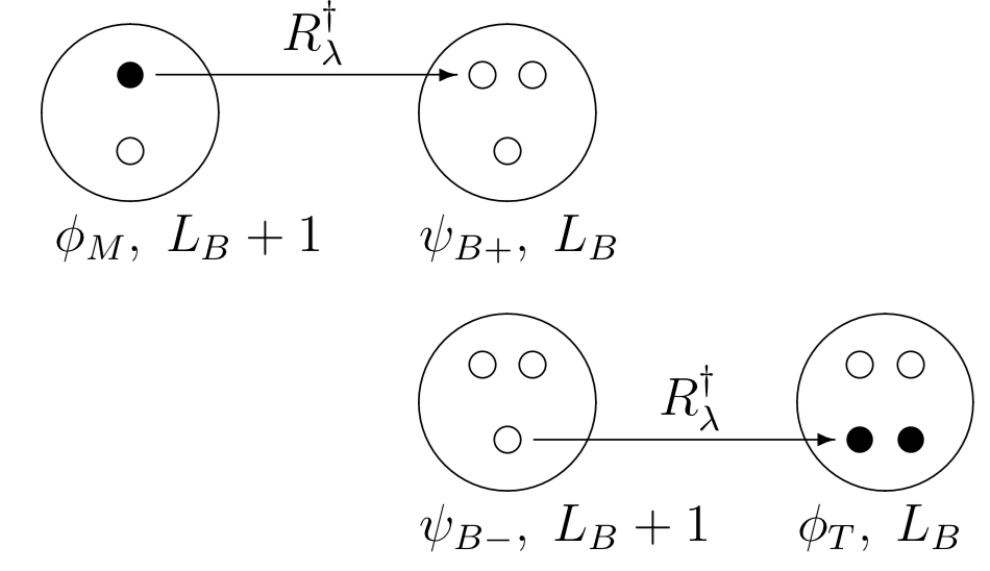} 
\caption{\lb{fig:MBTplet} The meson-baryon-tetraquark supersymmetric 4-plet $\{\phi_M, \phi_B^+, \phi_B^-, \phi_T\}$ follows from the two-step action of the supercharge operator $R^\dagger_\lambda$~\req{Rex}:   $\bar 3 \to 3 \times 3$ on the pion, followed by  $ 3 \to \bar 3 \times \bar 3$ on the negative chirality component of the nucleon.}
\end{figure}

Unfortunately, it is difficult to disentangle conventional hadronic quark states from exotic ones and, therefore, no clear-cut identification of tetraquarks for light hadrons, or hadrons with hidden charm or beauty, is possible~\cite{Brodsky:2016yod,Nielsen:2018uyn,Nielsen:2018ytt}. The situation is, however, more favorable for tetraquarks with open charm and beauty which may be stable under strong interactions and therefore easily identified~\cite{Karliner:2017qjm}. Our prediction~\cite{Dosch:2020hqm} for a doubly charmed stable boson $T_{cc}$ with a mass of 3870 MeV \ has been observed at LHCb a year later at 3875 MeV~\cite{LHCb:2021vvq}, and it is a member of the positive parity $J^P$ multiplet  $2^+,   \frac{3}{2}^+,   1^+$  of states  $\chi_{c2}(3565), ~ \Xi_{cc}(3770), ~ T_{cc}(3875)$.\footnote{For the inclusion of quark masses and longitudinal dynamics in holographic models see  Sec. 5.5.9  of Ref.~\cite{Gross:2022hyw}. In the limit of heavy quark masses HLFQCD leads to linear confinement, see Ref.~\cite{Trawinski:2014msa}.}
The occurrence of stable doubly beautiful tetraquarks and those with charm and beauty is also possible~\cite{Karliner:2017qjm, Dosch:2020hqm}.

\section{Color transparency}

A striking property of QCD phenomenology is color transparency (CT)~\cite{Brodsky:1988xz}, the reduced absorption of a hadron as it propagates through nuclear matter, if it is produced at high transverse momentum in a hard exclusive process, such as elastic lepton-proton scattering. The nuclear absorption reflects the effective size of the propagating hadron; {\it i.e.},  the separation between its color constituents.  CT has been confirmed in many experiments, such as semi-exclusive hard electroproduction, $e A \to e' \pi X$ for mesons produced at $Q^2 > 3 ~ {\rm GeV}^2$.  However, a recent JLab (Jefferson Laboratory) measurement for a proton electroproduced in carbon $e\, {\rm C}\to e' p X$ fails to observe CT at $Q^2$ up to 14.2 GeV$^2$~\cite{Bhetuwal:2020jes}. 

In a recent paper~\cite{Brodsky:2022bum}, the onset of CT for different hadrons was determined  by comparing the $Q^2$-dependence of the effective hadronic cross sections for the initial formation of a small color-singlet configuration, a pointlike configuration (PLC), using the generalized parton distributions (GPDs) from holographic light-front QCD~\cite{deTeramond:2018ecg}.  To discuss this problem, we can start from the flavor form factor of a hadron written in terms of its GPD, $H_q(x,t) \equiv H^q(x, \xi = 0, t)$  at zero skewness, $\xi$,  
\begin{align} \lb{Fqf}
F^q(t) & =   \int_0^1 dx \, H^q(x, t) \nn  \\
  & =  \int_0^1 dx \, q(x)  \exp \left[ t f(x)\right],  
\end{align}
where  $q(x)$ is the longitudinal parton distribution function (PDF) and $f(x)$ is the profile function. In HLFQCD $f(x)$ is flavor independent and, for hadron twist $\tau$, the~number of hadron constituents in a given Fock component, the  FF has the reparametrization-invariant integral representation,
expressed in terms of Euler’s Beta function~\cite{deTeramond:2018ecg}:
\begin{align}  \lb{FFB}
 F(t)_\tau &=  \frac{1}{N_\tau} B\left(\tau - 1, 1 - \alpha(t)\right) \nn \\   
 &= \frac{1}{N_\tau} \int_0^1 dx\, w'(x) w(x)^{-\alpha(t)} \left[1 - w(x) \right]^{\tau -2} ,
\end{align}
where $\alpha(t) = \alpha(0) + \alpha’ t$ is the Regge trajectory of the vector meson which couples to the quark current in the hadron, and~$N_\tau$ is a normalization factor.  The~trajectory $\alpha(t)$ can be computed  within the superconformal 
 LF holographic framework, and the intercept, $\alpha(0)$, incorporates the quark masses~\cite{deTeramond:2014asa, Dosch:2015nwa}. The~function $w(x)$ is a flavor-independent function with  $w(0) = 0, \, w(1) = 1$ and   $w'(x) \ge 0$. The~profile function $f(x)$ and the PDF 
$q_\tau(x)$ are determined by $w(x)$~\cite{deTeramond:2018ecg}:
\begin{align} 
f(x)&=\frac{1}{4\lambda}\log\Big(\frac{1}{w(x)}\Big) \lb{sigx} ,  \\ 
q_\tau(x)&=\frac{1}{N_\tau} w'(x)  w(x)^{-\alpha(0)} [1-w(x)]^{\tau-2} \lb{qx} ,
\end{align}
with $\alpha’ = 1 / 4 \lambda$. Boundary conditions  follow from the Regge behavior at $x \to 0$, $w(x) \sim x$,  and~ at $x \to 1$ from the inclusive-exclusive counting rule~\cite{Drell:1969km, Brodsky:1979qm}, $q_\tau(x) \sim (1-x)^{2 \tau - 3}$, which fix $w'(1) = 0 $. These physical conditions, together with the above-given constraints, basically determine the form of $w(x)$.

One expects from general considerations that the initial formation of a PLC  with a larger number of constituents --the proton for example, with its large phase space, has a lower probability to fluctuate to a small configuration as compared with a two-particle bound state, say the pion.  Consequently, it presents a larger transverse impact area as it traverses through the nucleus, and it will be slowed down or absorbed with greater probability as compared with a pion projectile with a smaller impact area for the same transverse momentum square $Q^2= - t$.  The onset of color transparency will be higher in $Q^2$ for the particle with a larger number of constituents compared to a hadron with fewer constituents.

The GPDs incorporate the far-off-shell components of the LFWF which controls the behavior of the  FF at large $Q^2$ and the power counting rules from the inclusive-exclusive connection~\cite{deTeramond:2018ecg}. The GPDs are thus essential to compute the $Q^2$ evolution of the effective transverse impact surface of a hadron from its physically observable value at $Q^2 = 0$, namely the hadron square radius, to its  high-virtuality PLC configuration. To this end we compute the transverse impact surface dependence on the momentum transfer, $t= - Q^2$, from the expectation value of the profile function $f(x)$\footnote{The procedure described here to compute the off-shell $Q^2$ evolution of the effective transverse size differs from the procedure in  reference~\cite{Burkardt:2003mb}, given instead in terms of relative impact variables.}:
\begin{align} \lb{a2tF}
\langle 4 f(t) \rangle_\tau  & =  
\frac{\int dx \, 4 f(x)  H^q_\tau(x,t)}{\int d x  H^q_\tau(x,t)}  \nn \\
&   =  \frac{1}{\lambda} \sum_{j = 1}^{\tau-1} \frac{1}{j - \alpha(t)},
\end{align}
which depends explicitly on the hadron's twist, $\tau$, and the properties of the specific quark current, which couples with the active quark in the hadron, characterized by the hadron's Regge trajectory,
$\alpha(t)$.  For large values of $ t = - Q^2$, Eq.~\req{a2tF} leads to
\begin{align} \lb{a2tasy}
\langle f(t) \rangle_\tau   \to  \frac{(\tau - 1)}{Q^2}.
\end{align}
It shows that, as expected, the effective transverse size decreases as $1/Q^2$, but the value of $Q^2$ required to contract all of the valence constituents to a color-singlet domain of a given transverse size grows as $\tau-1$, the number of the spectator constituents.

\begin{figure}[h]
\includegraphics[width=6.8cm]{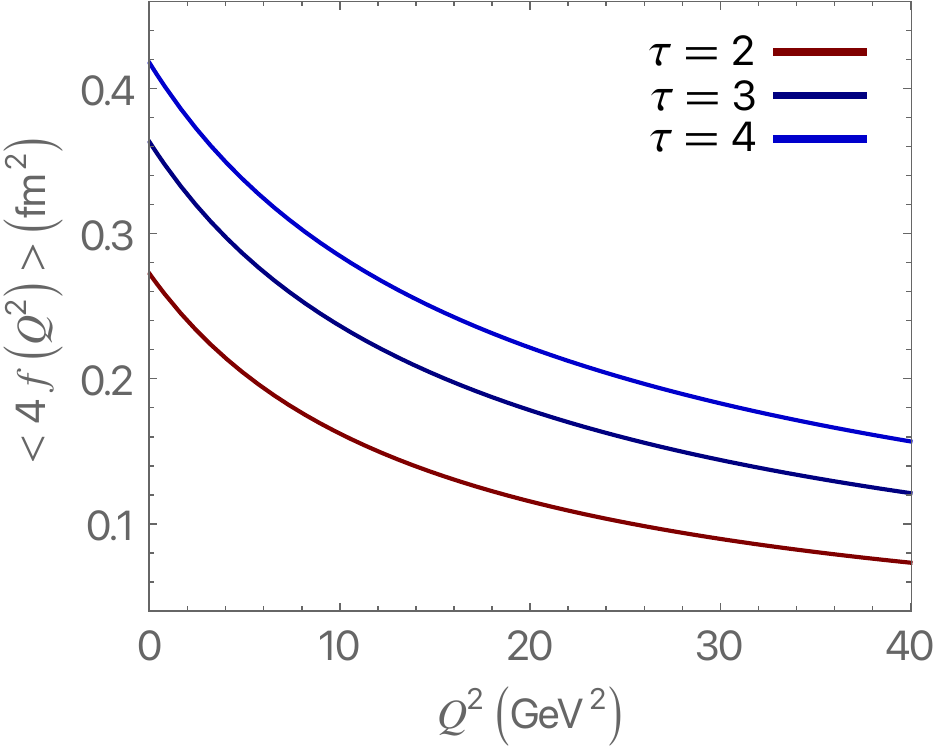} \hspace{10pt}
\includegraphics[width=7.2cm]{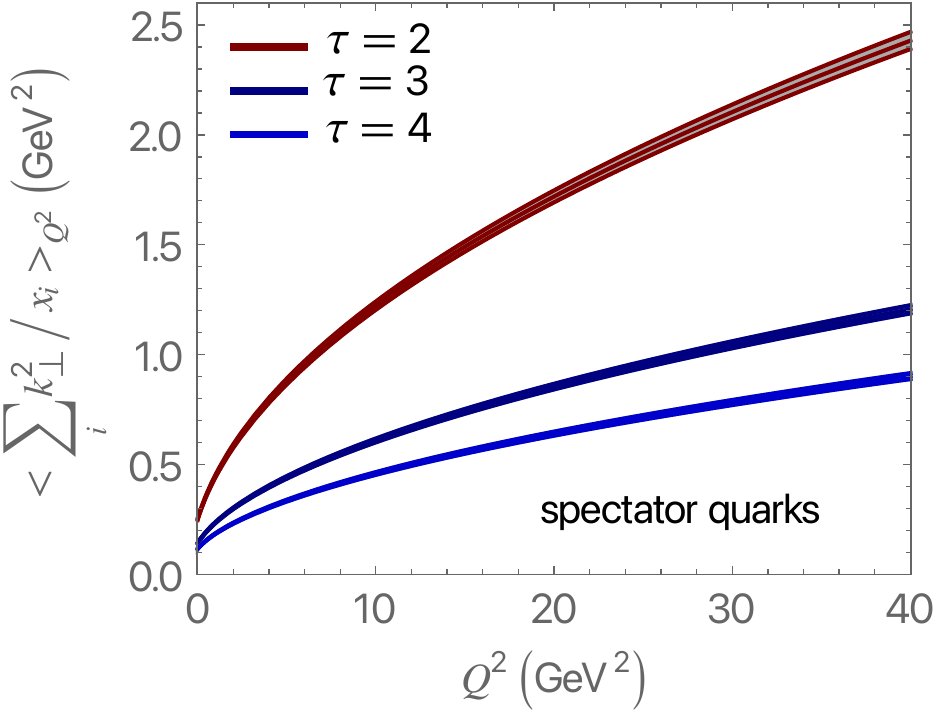}
\caption{The transverse impact area (left) $ \langle 4 \, f \left( t \right) \rangle$ as a function of $Q^2$ and the number of constituents  (twist), $\tau$, implies a significant delay in the onset of color transparency for $\tau > 2$. The off-shell $Q^2$ evolution (right) also shows an important dependence on the number of constituents.
 \lb{sigQ2}}
\end{figure}

We represent in Fig.~\ref{sigQ2} the critical dependence on the hadron's  twist $\tau$ of the effective transverse impact surface $ \langle 4 f(t) \rangle$.  It implies an important delay in the onset of CT in terms of the number of quark constituents. For example, the effective transverse impact surface for twist-2 at 4 GeV$^2$ is similar to that of twist-3 at 14 GeV$^2$.  Likewise, the impact surface at 4 GeV$^{2}$ for twist-2 is similar to that of twist-4 at 22 GeV$^2$, thus, implying an important delay in the onset of CT at intermediate energies in terms of the number of quark constituents. This is particularly relevant for the proton since it contains, at approximately equal probability, both twist-3, but also twist-4, in its LFWF.  The proton is thus expected to have a “two-stage” color transparency with the onset of CT differing for the spin-conserving (twist-3) Dirac form factor with a higher onset in $Q^2$ for the spin-flip Pauli (twist-4) form factor. This predicts a larger value of $Q^2$ in order to produce color transparency for the proton,  consistent  with the absence of proton CT in the present kinematic range of the JLab experiment~\cite{Bhetuwal:2020jes}. 

Complementary with the results in Ref.~\cite{Brodsky:2022bum}, we also show in Fig.~\ref{sigQ2} the PLC off-shell twist dependence computed from the $Q^2$ evolution of the invariant mass, along similar lines as the computation of the effective transverse surface in Ref.~\cite{Brodsky:2022bum}. The results in this figure clearly indicate that the pion goes off-shell much more rapidly than the proton. We only include in this figure the off-shell dependence of the spectator quarks, since the active quark remains close to its initial value at $Q^2 = 0$, independent of $Q^2$, indicating that most of the transverse momentum in this model is transferred to the spectator quarks.  A comparison of the expansion effects of the pion and proton formed in a high-momentum transfer collision was carried out in Ref.~\cite{Caplow-Munro:2021xwi} using superconformal baryon-meson symmetry in holographic QCD. No significant difference was found which could explain the JLab results in terms of the fast hadrons expanding their size, from the PLC configuration, as they escape the nucleus. Future experiments at higher values of $Q^2$ are critical to test the mechanisms discussed here which predict a significant twist-dependent delay in $Q^2$ for the onset of color transparency~\cite{Brodsky:1988xz}.

\section{Summary and outlook}

Although it has been half a century  since the $SU(3)_C$ color symmetry was introduced by Fritzsch and Gell-Mann~\cite{FGM} as the gauge field theory of the strong interactions, there has been little progress in understanding the fundamental physics which confines the colored quarks and gluons within hadrons. In fact, a basic understanding of the essential features of hadron physics from first-principles QCD has remained among the most important unsolved problems of the last 50 years in particle physics.  Hadronic characteristics are not explicit properties of the QCD Lagrangian but large distance emergent phenomena, where perturbative QCD, so successful in describing short distance phenomena, is not applicable.

Our present goal is trying to understand how emerging QCD properties would appear in an effective computational framework of hadron structure and as a step in this direction, we have reviewed in this article recent developments in hadron physics which follow from the application of superconformal quantum mechanics and light-front holography. This includes new insights into the physics of color confinement and the origin of the mass scale, chiral symmetry, the spectroscopy and dynamics of hadrons, the remarkable cancellation of quark kinetic energy and  the color confining potential in the pion, as well as surprising supersymmetric relations between the masses of mesons, baryons, and tetraquarks. We have also given a description of the holographic QCD approach to color transparency phenomena in nuclei~\cite{Brodsky:2022bum}.

There are other aspects and applications of HLFQCD which are not described in this article but are reviewed in reference~\cite{Gross:2022hyw}. For example,  LF holographic QCD also incorporates important elements for the study of hadron form factors, such as the connection between the hadron twist and the scaling behavior for large $Q^2$, and the incorporation of vector meson dominance which is relevant at lower energies. HLFQCD also incorporates features of pre QCD, such as the Veneziano model and Regge theory. Further extensions incorporate the exclusive-inclusive connection in QCD and provide nontrivial relations between hadron form factors and quark distributions~\cite{deTeramond:2018ecg, Liu:2019vsn}, including the intrinsic strange-antistrange~\cite{Sufian:2018cpj} and charm-anticharm~\cite{Sufian:2020coz} asymmetry distributions in the proton. HLFQCD has also been applied successfully to the description of gravitational form factors, the hadronic matrix elements of the energy momentum tensor, and gluon distributions in the proton and the pion~\cite{deTeramond:2021lxc}. Holographic QCD has also given new insights on the infrared behavior of the strong coupling in QCD~\cite{Deur:2022msf, Deur:2016tte}. Study of diffraction physics~\cite{Forshaw:2012im}  and the EMC effect in various nuclei~\cite{Kim:2022lng}, constitute other  examples of the application of the holographic light front ideas to QCD. More recently holographic QCD has been used to study the relation between the gluon density in a hadron and entanglement entropy to the high energy diffractive scattering behavior of hadrons~\cite{Dosch:2023bxj}. It was shown, for example, that the growth above the classical geometric cross section in proton-proton scattering at high energies is directly related to the increase of the internal quantum entropy from the entangled parton distribution in hadrons~\cite{Dosch:2023bxj}.

\section*{Acknowledgements}

We are grateful to Hans  G\"{u}nter Dosch for his invaluable contributions and to Alexandre Deur, Tianbo Liu, Raza Sabbir Sufian and Marina Nielsen for their collaboration on light-front holography and its implications. This work is supported by the Department of Energy, Contract DE--AC02--76SF00515.


\begin{thebibliography}{99}

\bibitem{FGM}
 H. Fritzsch and  M. Gell-Mann, 
 Current algebra: Quarks and what else?
 {\it Proceedings of the XIV International Conference on High Energy Physics, Chicago 1972,} Vol. 2, pp. 135–165 (1972)

\bibitem{Greenberg:1964pe}
O.~W.~Greenberg,
Spin and unitary spin independence in a paraquark model of baryons and mesons,
\href{https://journals.aps.org/prl/abstract/10.1103/PhysRevLett.13.598}{Phys. Rev. Lett. \textbf{13}, 598 (1964)}.

\bibitem{Han:1965pf}
M.~Y.~Han and Y.~Nambu,
Three triplet model with double $SU(3)$ symmetry,
\href{https://journals.aps.org/pr/abstract/10.1103/PhysRev.139.B1006}{Phys. Rev. \textbf{139}, B1006 (1965)}.

\bibitem{Bardeen:1972xk}
W.~A.~Bardeen, H.~Fritzsch and M.~Gell-Mann,
Light cone current algebra, $\pi^0$ decay, and $e^+ e^-$ annihilation,
{\it Scale and Conformal Symmetry in Hadron Physics,} Ed. by R. Gatto, John Wiley \& Sons, Inc. 1973, pp. 139--151 (1973)
[\href{https://arxiv.org/abs/hep-ph/0211388}{\tt arXiv:hep-ph/0211388}].

 \bibitem{Fritzsch:1973pi}
H.~Fritzsch, M.~Gell-Mann and H.~Leutwyler,
Advantages of the color octet gluon picture,
\href{https://www.sciencedirect.com/science/article/abs/pii/0370269373906254?via%3Dihub}{Phys. Lett. B \textbf{47}, 365 (1973)}.

\bibitem{Gross:2022hyw}
F.~Gross, E.~Klempt, S.~J.~Brodsky, A.~J.~Buras, V.~D.~Burkert, G.~Heinrich, K.~Jakobs, C.~A.~Meyer, K.~Orginos and M.~Strickland, \textit{et al.}
50 Years of quantum chromodynamics,
\href{https://link.springer.com/article/10.1140/epjc/s10052-023-11949-2}{Eur. Phys. J. C \textbf{83}, 1125 (2023)}
[\href{https://arxiv.org/abs/2212.11107}{\tt arXiv:2212.11107 [hep-ph]}].

\bibitem{Wilson:1974sk}
K.~G.~Wilson,
Confinement of quarks,
\href{https://journals.aps.org/prd/abstract/10.1103/PhysRevD.10.2445}{Phys. Rev. D \textbf{10}, 2445 (1974)}.

\bibitem{Maldacena:1997re}
 J.~M.~Maldacena,
 The large-$N$ limit of superconformal field theories and supergravity,
 \href{https://link.springer.com/article/10.1023%2FA%3A1026654312961}{Int.\ J.\ Theor.\ Phys.\  {\bf 38}, 1113 (1999)}
 [\href{http://arXiv.org/abs/hep-th/9711200}{\tt arXiv:hep-th/9711200}].
 
 \bibitem{Gubser:1998bc}
  S.~S.~Gubser, I.~R.~Klebanov and A.~M.~Polyakov,
  Gauge theory correlators from non-critical string theory,
  \href{http://www.sciencedirect.com/science/article/pii/S0370269398003773}{Phys.\ Lett.\ B {\bf 428}, 105 (1998)}
  [\href{http://arXiv.org/abs/hep-th/9802109}{\tt arXiv:hep-th/9802109}].

\bibitem{Witten:1998qj}
  E.~Witten,
  Anti-de Sitter space and holography,
  Adv.\ Theor.\ Math.\ Phys.\  {\bf 2}, 253 (1998)
  [\href{http://arXiv.org/abs/hep-th/9802150}{\tt arXiv:hep-th/9802150}].
  
\bibitem{Aharony:1999ti}
O.~Aharony, S.~S.~Gubser, J.~M.~Maldacena, H.~Ooguri and Y.~Oz,
Large $N$ field theories, string theory and gravity,
\href{https://www.sciencedirect.com/science/article/abs/pii/S0370157399000836?via%3Dihub}{Phys. Rept. \textbf{323}, 183 (2000)}
[\href{https://arxiv.org/abs/hep-th/9905111}{\tt arXiv:hep-th/9905111 [hep-th]]}.

 \bibitem {Dirac:1949cp}
  P.~A.~M.~Dirac,  
  Forms of relativistic dynamics,
  \href{https://doi.org/10.1103/RevModPhys.21.392}{Rev. Mod. Phys. \textbf{21}, 392 (1949).}

\bibitem{deTeramond:2008ht}
  G.~F.~de~T\'eramond and   S.~J. Brodsky,  
  Light-front holography: A first approximation to QCD, 
  \href{https://doi.org/10.1103/PhysRevLett.102.081601}{Phys. Rev. Lett. \textbf{102}, 081601 (2009)}
  [\href{https://arxiv.org/abs/0809.4899}{\tt arXiv:0809.4899 [hep-ph]}].
  
 \bibitem{Brodsky:2014yha}
  S.~J.~Brodsky,  G.~F.~de~T\'eramond,   H.~G.~Dosch, and  J.~Erlich,  
  Light-front holographic QCD and emerging confinement, 
  \href{https://doi.org/10.1016/j.physrep.2015.05.001}{Phys. Rept. \textbf{584}, 1 (2015)}
  [\href{https://arxiv.org/abs/1407.8131}{\tt arXiv:1407.8131 [hep-ph]}].
  
\bibitem{deTeramond:2018ecg} 
  G.~F.~de T\'eramond, T.~Liu, R.~S.~Sufian,  H.~G.~Dosch, S.~J.~Brodsky and A.~Deur [HLFHS],
  Universality of generalized parton distributions in light-front holographic QCD,
  \href{https://journals.aps.org/prl/abstract/10.1103/PhysRevLett.120.182001}{Phys.\ Rev.\ Lett.\  {\bf 120},  182001 (2018)}
  [\href{https://arxiv.org/abs/1801.09154}{\tt arXiv:1801.09154 [hep-ph]}].
  
 \bibitem{Liu:2019vsn} 
  T.~Liu, R.~S.~Sufian, G.~F.~de T\'eramond, H.~G.~Dosch, S.~J.~Brodsky and A.~Deur [HLFHS],
  Unified description of polarized and unpolarized quark distributions in the proton,
  \href{https://journals.aps.org/prl/abstract/10.1103/PhysRevLett.124.082003}{Phys.\ Rev.\ Lett.\  {\bf 124},  082003 (2020)}
  [\href{https://arxiv.org/abs/1909.13818}{\tt arXiv:1909.13818 [hep-ph]}].

\bibitem{deTeramond:2021lxc}
G.~F.~de T\'eramond, H.~G.~Dosch, T.~Liu, R.~S.~Sufian,   S.~J.~Brodsky and A.~Deur [HLFHS],
Gluon matter distribution in the proton and pion from extended holographic light-front QCD,
\href{https://journals.aps.org/prd/abstract/10.1103/PhysRevD.104.114005}{Phys. Rev. D \textbf{104}, 114005 (2021)}
[\href{https://arxiv.org/abs/2107.01231}{\tt arXiv:2107.01231 [hep-ph]}].
  
\bibitem{deAlfaro:1976vlx}
   V.~de~Alfaro,  S.~ Fubini,  and   G.~ Furlan,
   Conformal invariance in quantum mechanics,  
   \href{https://doi.org/10.1007/BF02785666}{Nuovo Cim. A \textbf{34}, 569 (1976).}
  
\bibitem{Fubini:1984hf}
   S.~Fubini and  E.~Rabinovici,  
   Superconformal quantum mechanics,  
   \href{https://doi.org/10.1016/0550-3213(84)90422-X}{Nucl. Phys. B \textbf{245}, 17 (1984).}

\bibitem{Akulov:1983hjq}
  V.~P.~Akulov  and   A.~I.~Pashnev,  
  Quantum superconformal model in $(1,2)$ space,  
  \href{https://doi.org/10.1007/BF01086252}{Theor. Math. Phys. \textbf{56}, 862 (1983).}  

\bibitem{Brodsky:2013ar}
  S.~J.~Brodsky,    G.~F.~de~T\'eramond,  and   H.~G.~Dosch,  
  Threefold complementary approach to holographic QCD,  
  \href{https://doi.org/10.1016/j.physletb.2013.12.044}{Phys. Lett. B \textbf{729}, 3 (2014)}
  [\href{https://arxiv.org/abs/1302.4105} {\tt arXiv:1302.4105 [hep-th]}].
  
\bibitem{deTeramond:2014asa}
  G.~F.~de~T\'eramond,  H.~G.~Dosch, and  S.~J.~Brodsky,  
  Baryon spectrum from superconformal quantum mechanics and its light-front holographic embedding,  
  \href{https://doi.org/10.1103/PhysRevD.91.045040} {Phys. Rev. D \textbf {91}, 045040 (2015)} 
  [\href{https://arxiv.org/abs/1411.5243} {\tt arXiv:1411.5243 [hep-ph]}].
  
\bibitem{Dosch:2015nwa}
  H.~G.~Dosch,   G.~F. de~T\'eramond, and  S.~J.~Brodsky, 
  Superconformal baryon-meson symmetry and light-front holographic QCD,
   \href{https://doi.org/10.1103/PhysRevD.91.085016}{Phys. Rev. D \textbf{91}, 085016 (2015)}
   [\href{https://arxiv.org/abs/1501.00959}{\tt arXiv:1501.00959 [hep-th]}].
     
\bibitem{Miyazawa:1966mfa}
   H.~Miyazawa,  
   Baryon number changing currents,  
   \href{https://doi.org/10.1143/PTP.36.1266}{Prog. Theor. Phys. \textbf{36}, 1266 (1966).}

\bibitem{Catto:1984wi}
   S.~Catto and F.~ Gursey,
   Algebraic treatment of effective supersymmetry, 
    \href{https://doi.org/10.1007/BF02902548}{Nuovo Cim. A  \textbf{86}, 201 (1985).}
  
\bibitem{Lichtenberg:1999sc}
  D.~B.~Lichtenberg,  
  Whither hadron supersymmetry?, in {\it International Conference on Orbis Scientiae 1999: Quantum Gravity,
  Generalized Theory of Gravitation and Superstring Theory Based Unification},
  28th Conference on High-Energy Physics and Cosmology  (1999) pp. 203--208, 
  \href{https://arxiv.org/abs/hep-ph/9912280}{\tt arXiv:hep-ph/9912280.}
  
\bibitem {Witten:1981nf}
  E.~Witten,  Dynamical breaking of supersymmetry,  
  \href{https://doi.org/10.1016/0550-3213(81)90006-7}{Nucl. Phys. B \textbf{188}, 513 (1981).}  
  
\bibitem{Brodsky:1988xz}
S.~J.~Brodsky and A.~H.~Mueller,
Using nuclei to probe hadronization in QCD,
\href{https://www.sciencedirect.com/science/article/abs/pii/0370269388907198?via%3Dihub}{Phys. Lett. B \textbf{206}, 685 (1988).}    

\bibitem{Brodsky:2022bum}
S.~J.~Brodsky and G.~F.~de T\'eramond,
Onset of color transparency in holographic light-front QCD,
\href{https://www.mdpi.com/2624-8174/4/2/42}{MDPI Physics \textbf{4}, 633 (2022)}
[\href{https://arxiv.org/abs/2202.13283}{\tt arXiv:2202.13283 [hep-ph]}].

\bibitem{Brodsky:1997de}
S.~J.~Brodsky, H.~C.~Pauli and S.~S.~Pinsky,
Quantum chromodynamics and other field theories on the light cone,
\href{https://www.sciencedirect.com/science/article/pii/S0370157397000896?via%3Dihub}{Phys. Rept. \textbf{301}, 299 (1998)}
[\href{https://arxiv.org/abs/hep-ph/9705477}{\tt arXiv:hep-ph/9705477 [hep-ph]}].

\bibitem{tHooft:1974pnl}
G.~'t Hooft,
A two-dimensional model for mesons,
\href{https://www.sciencedirect.com/science/article/abs/pii/0550321374900881?via%3Dihub}{Nucl. Phys. B \textbf{75}, 461 (1974)}.   

\bibitem{Pauli:1985pv}
H.~C.~Pauli and S.~J.~Brodsky,
Solving field theory in one space and one time dimension,
\href{https://journals.aps.org/prd/abstract/10.1103/PhysRevD.32.1993}{Phys. Rev. D \textbf{32}, 1993 (1985)}.

\bibitem{Hornbostel:1988fb}
K.~Hornbostel, S.~J.~Brodsky and H.~C.~Pauli,
Light-cone-quantized QCD in 1+1 dimensions,
\href{https://journals.aps.org/prd/abstract/10.1103/PhysRevD.41.3814}{Phys. Rev. D \textbf{41}, 3814 (1990)}.

\bibitem{Hornbostel:1988ne}
K.~Hornbostel,
The application of light cone quantization to quantum chromodynamics in one-plus-one dimensions,
Ph.~D. thesis Stanford University, 
\href{https://inspirehep.net/files/5382bfadf6016c08e11430938dd83262}{SLAC-PUB-0333 (1988)}.  

\bibitem{Gross:1973id} 
  D.~J.~Gross and F.~Wilczek,
  Ultraviolet behavior of non abelian gauge theories,
 \href{http://prl.aps.org/abstract/PRL/v30/i26/p1343_1}{Phys.\ Rev.\ Lett.\  {\bf 30}, 1343 (1973)}.
 
\bibitem{Politzer:1973fx} 
  H.~D.~Politzer,
  Reliable perturbative results for strong interactions?,
\href{http://prl.aps.org/abstract/PRL/v30/i26/p1346_1}{Phys.\ Rev.\ Lett.\  {\bf 30}, 1346 (1973)}.  

\bibitem{Polchinski:2001tt}
  J.~Polchinski and M.~J.~Strassler,
 Hard scattering and gauge/string duality,
  \href{http://prl.aps.org/abstract/PRL/v88/i3/e031601}{Phys.\ Rev.\ Lett.\  {\bf 88}, 031601 (2002)}
  [\href{http://arXiv.org/abs/hep-th/0109174}{\tt arXiv:hep-th/0109174}].
  
  \bibitem{Karch:2006pv}
  A.~Karch, E.~Katz, D.~T.~Son and M.~A.~Stephanov,
  Linear confinement and AdS/QCD,
 \href{http://prd.aps.org/abstract/PRD/v74/i1/e015005}{ Phys.\ Rev.\  D {\bf 74}, 015005 (2006)}
  [\href{http://arXiv.org/abs/hep-ph/0602229}{\tt arXiv:hep-ph/0602229}].
    
  \bibitem{deTeramond:2013it}
  G.~F. de~T\'eramond,  H.~G.~Dosch, and   S.~J.~Brodsky,  
  Kinematical and dynamical aspects of higher-spin bound-state equations in holographic QCD,
  \href{https://doi.org/10.1103/PhysRevD.87.075005}{Phys. Rev. D \textbf {87}, 075005 (2013)}
  [\href{https://arxiv.org/abs/1301.1651} {\tt arXiv:1301.1651 [hep-ph]}].
    
 \bibitem{Breitenlohner:1982jf}
 P.~Breitenlohner and D.~Z.~Freedman,
 Stability in gauged extended supergravity,
 \href{https://www.sciencedirect.com/science/article/abs/pii/0003491682901166?via%3Dihub}{Annals Phys. \textbf{144}, 249 (1982).}
   
  \bibitem{Kirsch:2006he} 
  I.~Kirsch,
  Spectroscopy of fermionic operators in AdS/CFT,
  \href{http://iopscience.iop.org/1126-6708/2006/09/052/}{JHEP\ {\bf 0609}, 052  (2006)}
  [\href{http://arXiv.org/abs/hep-th/0607205}{\tt arXiv:hep-th/0607205}].

   \bibitem{Abidin:2009hr}   
  Z.~Abidin and C.~E.~Carlson,
  Nucleon electromagnetic and gravitational form factors from holography,
 \href{http://prd.aps.org/abstract/PRD/v79/i11/e115003}{Phys.\ Rev.\  D {\bf 79}, 115003 (2009)}
  [\href{http://arXiv.org/abs/0903.4818}{\tt arXiv:0903.4818 [hep-ph]}]. 

 \bibitem{Gutsche:2011vb}
  T.~Gutsche, V.~E.~Lyubovitskij, I.~Schmidt and A.~Vega,
  Dilaton in a soft-wall holographic approach to mesons and baryons,
  \href{http://prd.aps.org/abstract/PRD/v85/i7/e076003}{ Phys.\ Rev.\ D {\bf 85}, 076003 (2012)}
  [\href{http://arXiv.org/abs/1108.0346}{\tt arXiv:1108.0346 [hep-ph]}].
  
 \bibitem{ParticleDataGroup:2022pth}
R.~L.~Workman \textit{et al.} [Particle Data Group],
Review of particle physics,
\href{https://doi.org/10.1093/ptep/ptac097}{PTEP  \textbf{2022}, 083C01 (2022).} 

\bibitem{Gao:2022ojh}
F.~Gao and M.~Yamada,
Determining wave equations in holographic QCD from Wilsonian renormalization group,
\href{https://journals.aps.org/prd/abstract/10.1103/PhysRevD.106.126003}{Phys. Rev. D \textbf{106}, 126003 (2022)}
[\href{https://arxiv.org/abs/2202.13699}{\tt arXiv:2202.13699 [hep-th]}].

\bibitem{Dosch:2016zdv}
H.~G.~Dosch, G.~F.~de T\'eramond and S.~J.~Brodsky,
Supersymmetry across the light and heavy-light hadronic spectrum II,
\href{https://journals.aps.org/prd/abstract/10.1103/PhysRevD.95.034016}{Phys. Rev. D \textbf{95},  034016 (2017)}
[\href{https://arxiv.org/abs/1612.02370}{\tt arXiv:1612.02370 [hep-ph]}].

 \bibitem{Brodsky:2016yod}
S.~J.~Brodsky, G.~F.~de T\'eramond, H.~G.~Dosch and C.~Lorc\'e,
Universal effective hadron dynamics from superconformal algebra,
\href{https://www.sciencedirect.com/science/article/pii/S0370269316302155?via%3Dihub}{Phys. Lett. B \textbf{759}, 171 (2016)}
[\href{https://arxiv.org/abs/1604.06746}{\tt arXiv:1604.06746 [hep-ph]}].

 \bibitem{Nielsen:2018ytt}
  M.~Nielsen,  S.~J.~Brodsky,  G.~F.~de~T\'eramond,  H.~G.~Dosch,  F.~S.~Navarra, and L.~ Zou, 
   Supersymmetry in the double-heavy hadronic spectrum,  
   \href{https://doi.org/10.1103/PhysRevD.98.034002}{Phys. Rev. D \textbf{98}, 034002 (2018)}
  [\href{https://arxiv.org/abs/1805.11567}{\tt arXiv:1805.11567 [hep-ph]}]. 
  
  \bibitem{Nielsen:2018uyn}
  M.~Nielsen and  S.~J.~Brodsky,  
  Hadronic superpartners from a superconformal and supersymmetric algebra, 
  \href{https://doi.org/10.1103/PhysRevD.97.114001}{Phys. Rev. D \textbf{97}, 114001 (2018)}
  [\href{https://arxiv.org/abs/1802.09652}{\tt arXiv:1802.09652 [hep-ph]}].

\bibitem{Karliner:2017qjm}
  M.~Karliner and  J.~L. Rosner,  Discovery of doubly-charmed $\Xi_{cc}$ baryon implies a stable $b b \bar{u} \bar{d}$ tetraquark,   
  \href{https://doi.org/10.1103/PhysRevLett.119.202001}{Phys. Rev. Lett \textbf{119}, 202001 (2017)}
  [\href{https://arxiv.org/abs/1707.07666} {\tt arXiv:1707.07666 [hep-ph]}].
 
\bibitem{Dosch:2020hqm}
   H.~G.~Dosch, S.~J.~Brodsky, G.~F.~de~T\'eramond, M.~Nielsen, and L.~ Zou,  
   Exotic states in a holographic theory,  
   \href{https://doi.org/10.1016/j.nuclphysbps.2021.05.035}{Nucl. Part. Phys. Proc. \textbf{312-317}, 135 (2021)} 
   [\href{https://arxiv.org/abs/2012.02496}{\tt arXiv:2012.02496 [hep-ph]}].
   
\bibitem{LHCb:2021vvq}
R.~Aaij \textit{et al.} [LHCb],
Observation of an exotic narrow doubly charmed tetraquark,
\href{https://doi.org/10.1038/s41567-022-01614-y}{Nature Phys. \textbf{18}, 751-754 (2022)}
[\href{https://arxiv.org/abs/2109.01038}{\tt arXiv:2109.01038 [hep-ex]}].      
   
\bibitem{Trawinski:2014msa}
A.~P.~Trawi\'nski, S.~D.~G\l{}azek, S.~J.~Brodsky, G.~F.~de T\'eramond and H.~G.~Dosch,
Effective confining potentials for QCD,
\href{https://journals.aps.org/prd/abstract/10.1103/PhysRevD.90.074017}{Phys. Rev. D \textbf{90}, 074017 (2014)}
[\href{https://arxiv.org/abs/1403.5651}{\tt arXiv:1403.5651 [hep-ph]}].

\bibitem{Bhetuwal:2020jes}
D.~Bhetuwal \textit{et al.} [Hall C],
Ruling out color transparency in quasielastic $^{12} {\rm C(e,e’p)}$ up to $Q^2$ of 14.2 (GeV/c)$^2$,
\href{https://journals.aps.org/prl/abstract/10.1103/PhysRevLett.126.082301}{Phys. Rev. Lett. \textbf{126}, 082301 (2021)}
[\href{https://arxiv.org/abs/2011.00703}{\tt arXiv:2011.00703 [nucl-ex]}].

 \bibitem{Drell:1969km}
  S.~D.~Drell and T.~M.~Yan,
 Connection of elastic electromagnetic nucleon form-factors at large $Q^2$
 and deep inelastic structure functions near threshold,
 \href{http://prl.aps.org/abstract/PRL/v24/i4/p181_1}{Phys.\ Rev.\ Lett.\  {\bf 24}, 181 (1970)}.
 
\bibitem{Brodsky:1979qm}
S.~J.~Brodsky and G.~P.~Lepage,
Exclusive processes and the exclusive-inclusive connection in quantum chromodynamics,
SLAC-PUB-2294 (1977). 

\bibitem{Burkardt:2003mb}
M.~Burkardt and G.~A.~Miller,
Color transparent general parton distributions,
\href{https://journals.aps.org/prd/abstract/10.1103/PhysRevD.74.034015}{Phys. Rev. D \textbf{74}, 034015 (2006)}
[\href{https://arxiv.org/abs/hep-ph/0312190}{\tt arXiv:hep-ph/0312190 [hep-ph]}].


\bibitem{Caplow-Munro:2021xwi}
O.~Caplow-Munro and G.~A.~Miller,
Color transparency and the proton form factor: Evidence for the Feynman mechanism,
\href{https://journals.aps.org/prc/abstract/10.1103/PhysRevC.104.L012201}{Phys. Rev. C \textbf{104}, L012201 (2021)}
[\href{https://arxiv.org/abs/2104.11168}{\tt arXiv:2104.11168 [nucl-th]}].

\bibitem{Sufian:2018cpj}
R.~S.~Sufian, T.~Liu, G.~F.~de T\'eramond, H.~G.~Dosch, S.~J.~Brodsky, A.~Deur, M.~T.~Islam and B.~Q.~Ma,
Nonperturbative strange-quark sea from lattice QCD, light-front holography, and meson-baryon fluctuation models,
\href{https://journals.aps.org/prd/abstract/10.1103/PhysRevD.98.114004}{Phys. Rev. D \textbf{98}, 114004 (2018)}
\href{https://arxiv.org/abs/1809.04975}{[\tt arXiv:1809.04975 [hep-ph]}].

\bibitem{Sufian:2020coz}
R.~S.~Sufian, T.~Liu, A.~Alexandru, S.~J.~Brodsky, G.~F.~de T\'eramond, H.~G.~Dosch, T.~Draper, K.-F.~Liu and Y.-B.~Yang,
Constraints on charm-anticharm asymmetry in the nucleon from lattice QCD,
\href{https://doi.org/10.1016/j.physletb.2020.135633}{Phys. Lett. B \textbf{808}, 135633 (2020)}
[\href{https://arxiv.org/abs/2003.01078}{\tt arXiv:2003.01078 [hep-lat]}].

\bibitem{Deur:2022msf}
A.~Deur, V.~Burkert, J.~P.~Chen and W.~Korsch,
Experimental determination of the QCD effective charge $\alpha_{g_1}(Q)$,
\href{https://www.mdpi.com/2571-712X/5/2/15}{Particles \textbf{5}, 171 (2022)}
[\href{https://arxiv.org/abs/2205.01169}{\tt arXiv:2205.01169 [hep-ph]}].

\bibitem{Deur:2016tte}
A.~Deur, S.~J.~Brodsky and G.~F.~de T\'eramond,
The QCD running coupling,
\href{https://www.sciencedirect.com/science/article/abs/pii/S0146641016300035?via%3Dihub}{Nucl. Phys. \textbf{90}, 1 (2016)}
[\href{https://arxiv.org/abs/1604.08082}{\tt arXiv:1604.08082 [hep-ph]}].

\bibitem{Forshaw:2012im}
J.~R.~Forshaw and R.~Sandapen,
An AdS/QCD holographic wave function for the $\rho$ meson and diffractive $\rho$ meson electroproduction,
\href{https://journals.aps.org/prl/abstract/10.1103/PhysRevLett.109.081601}{Phys. Rev. Lett. \textbf{109}, 081601 (2012)}
[\href{https://arxiv.org/abs/1203.6088}{\tt arXiv:1203.6088 [hep-ph]}].

\bibitem{Kim:2022lng}
D.~N.~Kim and G.~A.~Miller,
Light-front holography model of the EMC effect,
\href{https://journals.aps.org/prc/abstract/10.1103/PhysRevC.106.055202}{Phys. Rev. C \textbf{106},  055202 (2022)}
[\href{https://arxiv.org/abs/2209.13753}{\tt arXiv:2209.13753 [nucl-th]}].

\bibitem{Dosch:2023bxj}
H.~G.~Dosch, G.~F.~de T\'eramond and S.~J.~Brodsky,
Entropy from entangled parton states and high-energy scattering behavior,
\href{https://www.sciencedirect.com/science/article/pii/S0370269324000790?via%3Dihub}{Phys. Lett. B \textbf{850}, 138521 (2024)}
[\href{https://arxiv.org/abs/2304.14207}{\tt arXiv:2304.14207 [hep-ph]}].


\end{thebibliography}
\end{document}